# Analysis of the detective quantum efficiency of a neutron image plate detector

S. Masalovich [a,b]

[a] Heinz Maier-Leibniz Zentrum (MLZ), Technische Universität München, Lichtenbergstr. 1, 85748 Garching, Germany
[b] Jülich Centre for Neutron Science, Forschungszentrum Jülich GmbH, 52425 Jülich, Germany
*Sergey.Masalovich@frm2.tum.de*

**Abstract**

The aim of the present note is to derive a general formula for a detective quantum efficiency (DQE) of a neutron image plate detector which includes all essential parameters of the system. Relying on that formula, the effects of a variety of processes involved in the formation of an output signal are discussed and estimated. The theoretical value of the DQE is compared with the experimental one measured for the commercial neutron image plate BAS-IP ND 2025 (FUJIFILM).

## 1. Introduction

In neutron scattering experiments where the timing is not required (e.g. static neutron diffraction and imaging) a counting detector may be replaced with an integrating detector, such as a neutron image plate [1 - 4].

An image plate is a two-dimensional position sensitive integrating detector that was initially developed for X-ray imaging [5]. It operates in many ways similar to a common photo-film where a latent image is first formed by exposure of a film to a radiation and then a visible image is restored by a post-processing. An image plate for X-rays constitutes a thin layer ($\sim 100$ μm) of a fine powder of the storage phosphor $BaFBr:Eu^{2+}$ in organic binder coated on a flexible supporting foil. The distinctive properties of an image plate are: a) high efficiency, b) a very high spatial resolution (some tens of micrometers), c) an arbitrarily large active area, d) a large dynamical range (liner response for up to seven orders of magnitude for recorded signal) and e) it is re-usable after short erasing. Taken together these properties make an image plate a very attractive detector in many research areas with X-ray imaging and diffraction [6].

By adding a fine powder of a neutron absorber to phosphor crystals one can build a neutron sensitive image plate [7, 8]. At present, $Gd_2O_3$ is used as a neutron absorber in commercially available neutron image plates [9]. The neutron absorption is accompanied by ionizing radiation that gives rise to creation of free electrons and holes in the storage phosphor crystals. The electrons are trapped at halide vacancies forming different types of F-centers while the holes are captured at $Eu^{2+}$ ions resulting in Eu-hole complexes creation [10]. The concentration of the F-centers and Eu-hole complexes is proportional to the locally deposited energy so that the latent image of the incident flux is formed. The image recovering is commonly performed by sequential point-by-point scanning of the image plate surface by a focused laser beam. During this readout process the laser light excites F-center electrons, which in turn recombine with nearby holes resulting in the characteristic blue $Eu^{2+}$-luminescence called Photo-Stimulated Luminescence (PSL) which is detected by a photo-detector. The temporal sequence of the PSL signals measured synchronously with the laser beam movement provides a way for 2D imaging.

The information in an image plate is accumulated over an exposure time and the signals from all incident neutrons add up to give a final image. Because every measured PSL value represents a joint result of many stochastic processes, the distribution of the final measured signals is subject to the law of probability. As a consequence, the summed signal in each element of a picture has statistical fluctuations which constitute a noise. So, the signal-to-noise ratio in an image is to be considered as one of the main parameters that should characterize the detector. A quantitative measure of that parameter can be estimated with the use of the value of a Detective Quantum Efficiency (DQE) (see, e.g., [11]) defined as

$$DQE = \frac{\left(\bar{n}_{out}/\sigma_{out}\right)^2}{\left(\bar{n}_{in}/\sigma_{in}\right)^2} \quad (1)$$

Here $\bar{n}_{out}$ is the mean value of the measured output PSL signal for a given element of an image, $\sigma_{out}$ is the standard deviation of that value. Analogously, $\bar{n}_{in}$ and $\sigma_{in}$ are the mean value and the standard deviation of the input signal (incident neutron flux) for the same element of an image. Hence in order to estimate the DQE of the detector it is necessary to calculate the statistical fluctuations of the output signals with the use of a prior knowledge of the elementary processes involved in the formation of the signal.

The aim of the present work is to derive a general formula for the DQE of a neutron image plate detector which include all essential parameters of the system. Relying on that formula, the main requirements for the detector are derived and analyzed.

## 2. Image formation as a cascade process

Fig.1 shows the physical processes involved in the formation of a latent neutron image and successive recovering of it with the use of a laser beam. As can be seen, the sequence of the processes comprises a cascade in which an output of every single process constitutes an input signal for a successive process. Each process in the cascade represents some stochastic process which contributes to the total fluctuation of the final output signal and thus affects the signal-to-noise ratio. By calculating the DQE (see Eq.(1)), one can estimate on how much greater are statistical fluctuations of the output signals as compared with statistical fluctuations of the input signals. In the case of a neutron image plate detector the input signal is a random number of neutrons arriving at a unit area of an image plate over an exposure time. Since the measured signal is an electrical signal detected at the output of a photo-detector (in general, a photomultiplier tube (PMT)) then as a final output signal we take a number of electrons arriving at the anode of a PMT.

The statistical properties of a cascade process can be determined by application of the method of a generating function [12, 13] defined for a distribution of random variables with integer values. Let a variable $k$ take random discrete values 0, 1, 2 ... with the probabilities $p_0$, $p_1$,

$p_2$ ... The probability generating function for this variable is defined as

$$\varphi(s) = \sum_{k=0}^{\infty} p_k s^k \ , \qquad (2)$$

where the arbitrary parameter $s$ must satisfy the condition $0 < s \leq 1$ to ensure the convergence of the series in Eq.(2). The function $\varphi(s)$ completely defines the statistical properties of the variable $k$ since it contains all probabilities $p_k$ as coefficients in a power expansion. It is obvious that $\varphi(1) = 1$. The mean value and variance of the variable $k$ are derived from $\varphi(s)$ as follows:

$$\bar{k} = \varphi'(1) \ , \qquad (3)$$

$$D(k) = \varphi''(1) + \varphi'(1) - [\varphi'(1)]^2 \ , \qquad (4)$$

where we used $\varphi'(s) \equiv \dfrac{d\varphi(s)}{ds}$ and $\varphi''(s) \equiv \dfrac{d^2\varphi(s)}{ds^2}$.

Let the statistical properties of the first process in a cascade be described with the probability generating function $\varphi_1(s)$. We can further define the probability generating function $\varphi_{2,1}(s)$ that describes the statistical properties of the second process as a response on every single input signal (one neutron, one photon, one electron and so on) emerging at the output of the first process. Hence by the property of a generating function the output signal distribution will be described by the composition of functions

$$\varphi_2(s) = \varphi_1(\varphi_{2,1}(s)) \ . \qquad (5)$$

In the case of an image plate, the first process is described by Poisson distribution with the generating function

$$\varphi(s) = e^{-N_0 + N_0 s} \qquad (6)$$

with $N_0$ being the mean number of the arrived neutrons over the exposure period. The probability distribution of the number of neutrons absorbed in the detector per one incident neutron (obviously 0 or 1) is described by the binomial law with the probability generating function

$$\varphi_{2,1}(s) = q + ps \ , \qquad (7)$$

where $p$ is the probability for a neutron to be absorbed, and $q = 1 - p$. Substitution of Eqs. (6) and (7) into Eq.(5) gives:

$$\varphi_2(s) = e^{-N_0 + N_0(q + ps)} = e^{-pN_0 + pN_0 s} \ . \qquad (8)$$

Hence the number of absorbed neutrons also adheres to Poisson statistics with the well known formulas for the mean value $\bar{k} = pN_0$ and the variance $D(k) = pN_0$. We will use this in the following evaluation.

Analogously, for other elementary processes in a cascade one can write

$$\varphi_{n+1}(s) = \varphi_n(\varphi_{n+1,1}(s)) ,\qquad(9)$$

where subindex "n+1,1" indicates the probability generating function $\varphi_{n+1,1}(s)$ of the (n+1)-th process. By applying Eq.(9) successively for all processes in a cascade one can obtain the generating function for the entire cascade. Eqs.(3) and (4) then give the parameters of the final statistical distribution. In a general case of an N-step-cascade one can develop

$$\bar{n} = \bar{k}_1 \bar{k}_2 ... \bar{k}_N ,\qquad(10)$$

$$D(n) = D(k_1)\bar{k}_2^2 \bar{k}_3^2 ... \bar{k}_N^2 + \bar{k}_1 D(k_2)\bar{k}_3^2 \bar{k}_4^2 ... \bar{k}_N^2 + \bar{k}_1 \bar{k}_2 D(k_3)\bar{k}_4^2 \bar{k}_5^2 ... \bar{k}_N^2 + ... + \bar{k}_1 \bar{k}_2 ... \bar{k}_{N-1} D(k_N) ,\qquad(11)$$

where $k_i$ refers to a random number at the output of the $i$-th process per a unit input signal (i > 1), $\bar{k}_i$ and $D(k_i)$ are the mean value and the variance of the $k_i$. By definition, the relative fluctuation of the variable $n$ is

$$\delta(n) = \frac{\sqrt{D(n)}}{\bar{n}} .\qquad(12)$$

Substituting Eq.(10) and (11) into Eq.(12) one obtains

$$\delta^2(n) = \delta^2(k_1) + \frac{1}{\bar{k}_1}\delta^2(k_2) + \frac{1}{\bar{k}_1 \bar{k}_2}\delta^2(k_3) + ... + \frac{1}{\bar{k}_1 \bar{k}_2 ... \bar{k}_{N-1}}\delta^2(k_N) .\qquad(13)$$

## 3. DQE of a neutron image plate detector

We now apply the method of a generating function to study the statistical properties of a neutron image plate detector. The cascade process in this case is shown in Fig.1. Below we identify all elementary processes in that cascade and consider their statistical properties.
1. A random number of neutrons ($k_1$) are arriving at the specified area of the image plate over the exposure time. This process is described by Poisson distribution with the mean value $\bar{k}_1$ and the variance $D(k_1) = \bar{k}_1$.
2. An absorption of a neutron is described by the binominal distribution where the variable $k_2$ can take only two values: 0 and 1. If we denote the neutron absorption efficiency by $\eta$, then one can derive $\bar{k}_2 = \eta$ and $D(k_2) = \eta(1-\eta)$
3. Every neutron absorbed in the detector gives rise to ionizing radiation and the eventual creation of trapped electrons (F-centers) and holes in phosphor crystals. But only a part of them are photostimulable and we will consider just these centers. The number of such F-centers is a random number which fluctuation is essentially governed by the fluctuation of the value of the energy deposited in the crystals. Since the phosphor crystals are contained in a mixture with other materials (crystals of neutron absorber and binder), the distribution of an absorbed energy between them is random. In addition, the reaction of neutron absorption in Gd has many channels with a large fluctuation of released energy [14 - 16]. Let $k_3$ be the number of photostimulable F-centers created per one absorbed neutron. We define the mean value as $\bar{k}_3 = N_F$ and the variance $D(k_3) = D(N_F)$.

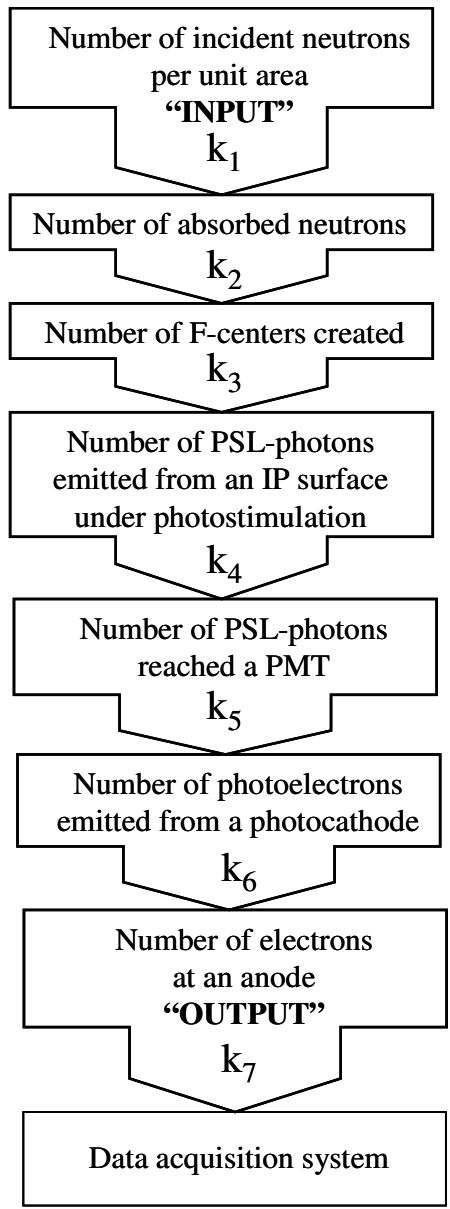

Fig.1 Sequence of processes

4. The electrons trapped in F-centers and holes captured at $Eu^{2+}$ ions form a latent image. The reading of the image is commonly performed by a laser light (so-called photo-stimulation). Two main stochastic processes related to the photo-stimulation are: a) creation of a PSL photon within the volume of an image plate and b) escape of the PSL photon out of an image plate. For the sake of simplicity we consider these two processes as a single process that leads to the emission of PSL photons. Let the variable $k_4$ describes the number of PSL photons emitted by the image plate per one F-center. We define the mean value and the variance of that distribution as $\bar{k}_4$ and $D(k_4)$.

5. The emitted PSL photons are guided to a photo-detector (PMT) with the use of a light-guide. The number of photons $k_5$ arriving at the detector per one photon emitted by the image plate can also take values 0 and 1. If we denote the guiding efficiency as $\bar{k}_5 = p$, then the variance can be found with the formula: $D(k_5) = p(1-p)$.

6. The PSL photon incident upon the photocathode of a PMT creates free photo-electrons. We assume that the number of electrons per one incident photon may only take values 0 and 1 (the maximal quantum efficiency of a commercial photocathode is about 0.25 for given PSL photons of energy $E_{ph} = 3.1 eV$ [17]). The mean value and variance of that distribution are denoted by $\bar{k}_6 = f$ and $D(k_6) = f(1-f)$.

7. Each photo-electron through the multiplication in a PMT leads to the creation of an electron cloud at an anode. Let $k_7$ be the number of electrons arriving at the anode per one electron emitted by the photocathode. The mean value of that number is called a multiplication factor: $\bar{k}_7 = M$. The variance of that number is denoted by $D(k_7) = D(M)$.

We consider the number of electrons at an anode to be the measured signal for a given element of an image plate. Substituting the statistical parameters of the processes mentioned above into Eqs. (10) and (11) we arrive at

$$\bar{n}_{out} = \bar{k}_1 \bar{k}_2 ... \bar{k}_7 ,\qquad(14)$$

$$\frac{D(n_{out})}{\bar{n}_{out}^2} = \frac{D(k_1)}{\bar{k}_1^2} + \frac{1}{\bar{k}_1}\frac{D(k_2)}{\bar{k}_2^2} + \frac{1}{\bar{k}_1\bar{k}_2}\frac{D(k_3)}{\bar{k}_3^2} + ... + \frac{1}{\bar{k}_1\bar{k}_2...\bar{k}_6}\frac{D(k_7)}{\bar{k}_7^2} .\qquad(15)$$

For the signal at the input one can write

$$\bar{n}_{in} \equiv \bar{k}_1 \ , \tag{16}$$

$$D(n_{in}) \equiv D(k_1) = \bar{k}_1 \ . \tag{17}$$

The DQE of an image plate detector is now obtained by substituting Eqs. (14-17) into Eq.(1):

$$DQE = \frac{\eta}{1 + \frac{D(N_F)}{\bar{N}_F^2} + \frac{1}{\bar{N}_F} \cdot \frac{D(k_4)}{\bar{k}_4^2} + \frac{1}{\bar{N}_F \bar{k}_4 pf}\left(1 - pf + \frac{D(M)}{M^2}\right)} \ . \tag{18}$$

This formula only slightly differs from the one presented in [18].

**4. Analysis of the DQE and a possible optimization of the detector**

The performance of an image plate detector in terms of statistical fluctuations can now be evaluated with the use of Eq.(18). This equation contains all principal parameters that describe both the efficiency of an image plate and the efficiency of a reading setup (a laser beam scanner and a photo-detector). The first obvious result is that the DQE is always less than the neutron absorption efficiency $\eta$. Furthermore, if we assume $N_F \gg 1$ (which is common [18 - 20]) then it follows that the DQE value is primarily determined by the neutron absorption efficiency and by the relative fluctuation of the energy deposited in phosphor crystals. It is significant that the efficiency of the readout process appears in the expression for the DQE only with the factor $1/\bar{N}_F$. Thus its contribution to the whole efficiency can be neglected unless the values $\bar{k}_4$, $p$ and $f$ are made extremely small (which, of course, not the case in a common practice). So, in order to optimize the performance of an image plate detector the condition

$$\frac{\eta}{1 + \frac{D(N_F)}{\bar{N}_F^2}} \to \max \tag{19}$$

should be fulfilled. In particular, it means that phosphor and neutron absorber crystals should comprise a very homogeneous mixture of fine powders so that the second term in the denominator of Eq.(19) tends to its minima. However, as noted earlier, the fluctuation of the energy deposited in the phosphor crystals is determined not only by the homogeneity of the powder mixture but also by the fluctuation of the released energy itself. Actually, the reaction of neutron absorption by Gd is a multichannel reaction with emission of a broad spectrum of γ-rays and electrons (internal conversion (IC) electrons and Auger electrons) [14, 21] thus leading to an inherent fluctuation of the energy output. In addition, the two isotopes, $^{155}$Gd and $^{157}$Gd, responsible for the neutron absorption in natural Gd, feature different spectra of ionizing radiation which may lead to increased energy fluctuation. Thus the question of the need to use isotope enrichment should be also raised.

With the aim to make an estimate of the term $D(N_F)/\bar{N}_F^2$ in Eq.(19) we only consider the IC-electrons since they constitute the main source of the energy deposited in the phosphor crystals. Table 1 presents the spectra of conversion electrons emitted by different isotopes of Gd after neutron absorption [15]. The mean values and the variances of the released energy for these isotopes are calculated and shown in the lower part of Table 1.

Table 1: Spectra of IC-electrons in (n,Gd) reaction

| $^{156}$Gd IC-electrons | | $^{158}$Gd IC-electrons | |
|---|---|---|---|
| Energy (keV) | Intensity/neutron | Energy (keV) | Intensity/neutron |
| 0 (no electrons) | 0.383 | 0 (no electrons) | 0.4044 |
| 39 | 0.224 | 29 | 0.121 |
| 81 | 0.266 | 71 | 0.330 |
| 88 | 0.062 | 78 | 0.076 |
| 149 | 0.045 | 131 | 0.042 |
| 191 | 0.016 | 173 | 0.018 |
| 198 | 0.003 | 180 | 0.0038 |
| 246 | 0.001 | 228 | 0.0048 |
| Electron energy per captured neutron | | Electron energy per captured neutron | |
| mean value $\mu_5$ = 46.34 | variance $D_5$ = 2179.6 | mean value $\mu_7$ = 59.27 | variance $D_7$ = 3972.9 |

For the natural mixture of isotopes the mean value $\mu$ and the variance $D$ of the IC-electron energy are

$$\mu = q_5 \mu_5 + q_7 \mu_7, \qquad (20)$$

$$D = q_5 D_5 + q_7 D_7 + q_5 q_7 (\mu_5 - \mu_7)^2, \qquad (21)$$

where $q_5$ and $q_7$ are the probabilities that the captured neutron will be absorbed by $^{155}$Gd or $^{157}$Gd respectively. With the known absorption cross-section [22] and abundance for each isotope one finds $q_5 = 0.182$ and $q_7 = 0.818$. Consequently, for the $^{nat}$Gd we obtain $\mu = 56.9$ keV and $D = 3671.4$ keV$^2$. When this result is compared with that of $^{157}$Gd (see Table 1), it becomes clear that the fluctuations in a natural gadolinium are strongly dominated by $^{157}$Gd and thus the isotope enrichment will not decrease the fluctuations.

The calculated relative variance for the $^{nat}$Gd is $\frac{D}{\mu^2} = 1.13$. So, the maximal DQE value (see Eq.(19)) may be estimated as $DQE_{th} \approx 0.47\eta$. Since for slow neutrons the absorption efficiency $\eta$ usually is very close to unity, even for the very thin layer of gadolinium, we arrive to the following estimate: $DQE_{th} \approx 0.47$. This theoretical value is in a good agreement with the experimental value $DQE_{exp} = 0.61$ measured[1] for the commercial image plate BAS IP ND 2025 (FUJI) exposed to neutrons with the wavelength λ = 2.34Å [23]. The slightly higher number for the measured $DQE_{exp}$ in comparison with the calculated one can be partly ascribed to the contribution of Auger electrons and γ-rays in the (n,Gd) reaction. One can expect, however, the contribution from Auger electrons to be small since these electrons are characterized by low energies and thus short ranges in Gd$_2$O$_3$ crystals and in a binder. As for the γ-rays, they are mostly of high energy and the energy loss in the phosphor crystals can be estimated with the use of the photon energy stopping power of about 0.02 keV/μm [23]. So, in case of a thin phosphor layer (about 50 μm in the used image plate) the energy deposition by γ-rays is of little importance. More appropriate explanation of the greater value for the $DQE_{exp}$ is a narrowing of the spectrum of the energy deposited by IC-electrons caused by the greater absorption of low-energy electrons in Gd crystals and a binder. Anyway, it is seen that the calculated $DQE_{th}$ makes a good estimate of the signal-to-noise ratio of the detector.

---

[1] The recent revision revealed that the value 0.78 published for the measured DQE in [23] is actually not the DQE but the square root of the measured DQE. Thus the correct value is $0.78^2 = 0.61$.

## 5. Summary

For a neutron image plate detector the process of an image creation and subsequent readout is analyzed. Considering the process as a cascade of successive events, the statistical fluctuations of the output signal is estimated and the formula for the Detective Quantum Efficiency (DQE) is derived. The theoretical value of the DQE is compared with the experimental value measured for the commercial neutron image plate BAS IP ND 2025 and the agreement is discussed. Despite the good performance of the measured image plate, it should be noted that its application is limited to measurements where a γ-ray background is rather low. Therefore an experimental study on the sensitivity of a commercial neutron image plate to γ-ray background at a given neutron instrument is required. If the sensitivity is notable, then image plates with other neutron convertors like $^6$Li [1, 9, 23, 24] or $^{10}$B [18] need to be built and optimized [25].

Finally, it is worth noting that Eq.(18) can be readily applied to estimate the signal-to-noise ratio for a X-ray image plate by the use of appropriate parameter values.

**Acknowledgements**

The author is deeply grateful to Alexander Ioffe and Thomas Brückel (Jülich Centre for Neutron Science) for their valuable comments and suggestions.